\documentclass[sigconf,screen]{acmart}

\usepackage{color}      
\usepackage{xspace}
\usepackage{enumitem}
\usepackage{soul}

\usepackage{wrapfig}
\usepackage{graphicx}

\AtBeginDocument{%
  \providecommand\BibTeX{{%
    \normalfont B\kern-0.5em{\scshape i\kern-0.25em b}\kern-0.8em\TeX}}}

\begin{document}

\title{\Polyglot: A Personalized and Gamified eTutoring System}

\author{Antonio Bucchiarone}
\affiliation{%
  \institution{Fondazione Bruno Kessler}
  \city{Trento}
  \country{Italy}
}
\email{bucchiarone@fbk.eu}

\author{Tommaso Martorella}
\affiliation{%
  \institution{Università di Pisa}
  \country{Pisa, Italy}
}
\email{tom.martorella@gmail.com}

\author{Diego Colombo}
\affiliation{%
  \institution{Microsoft Corporation Redmond}
  \city{WA, United States}
  \country{}}
 \email{diego.colombo@microsoft.com}

\renewcommand{\shortauthors}{Antonio Bucchiarone, Tommaso Martorella, and Diego Colombo}

\newcommand{\code}[1] {{\small\sffamily #1}}
\newcommand{\Polyglot}{PolyGloT}
\newcommand{\codesmall}[1] {{\footnotesize\sffamily #1}}

\begin{abstract}
The digital age is changing the role of educators and pushing for a paradigm shift in the education system as a whole. Growing demand for general and specialized education inside and outside classrooms is at the heart of this rising trend. In modern, heterogeneous learning environments, the one-size-fits-all approach is proven to be fundamentally flawed. Individualization through adaptivity is, therefore, crucial to nurture individual potential and address accessibility needs and neurodiversity. By formalizing a learning framework that takes into account all these different aspects, we aim to define and implement an open, content-agnostic, and extensible eTutoring platform to design and consume adaptive and gamified learning experiences. Adaptive technology supplementing teaching can extend the reach of every teacher, making it possible to scale 1-1 learning experiences. There are many successful existing technologies available but they come with fixed environments that are not always suitable for the targeted audiences of the course material. This paper presents \Polyglot, a system able to help teachers to design and implement a gamified and adaptive learning paths. Through it we address some important issues including the engagement, fairness, and effectiveness of learning environments. We do not only propose an innovative platform that could foster the learning process of different disciplines, but it could also help teachers and instructors in organizing learning material in an easy-access repository.
\end{abstract}



\maketitle

\section{Introduction}
\label{sec:introduction}
\textbf{Adaptive learning} is the delivery of personalized learning experiences that address an individual's unique needs instead of a one-size-fits-all approach. It may be achieved through just-in-time feedback, personalized learning paths, ad-hoc resources, or another wide array of techniques. But why is it important? Education is universally recognized as one of the factors with the highest impact on society and the individual. The United Nations included education in their 2030 Agenda for Sustainable Development\footnote{\url{https://www.un.org/sustainabledevelopment/education/}}. UNESCO launched the Global Education Coalition\footnote{\url{https://en.unesco.org/covid19/educationresponse/globalcoalition}} in response to the COVID-19 pandemic. The European Union created the Digital Education Action Plan (2021-2027)\footnote{\url{https://education.ec.europa.eu/focus-topics/digital-education/digital-education-action-plan}} to foster and support the adaptation of educational systems in the digital age. This collective global effort is motivated by a continuously increasing technology availability and a rising global enrolment rate. Furthermore, according to UNESCO, higher education is the fastest-growing sector\footnote{Data from UNESCO \cite{UNESCO_FLP} and Statista (retrieved on 29 June 2022). \\
\url{https://www.statista.com/statistics/1226999/net-enrollment-rate-in-primary-school-worldwide/} and \url{https://www.statista.com/statistics/1227022/net-enrollment-rate-in-secondary-school-worldwide/}}, with its global enrolment rate doubled in the last twenty years \cite{UNESCO_FLP}.

Rising trends such as \textbf{flexible learning pathways} and \textbf{micro-credentials}\footnote{ \url{https://education.ec.europa.eu/education-levels/higher-education/micro-credentials}} tend toward more versatile forms of content delivery and credential recognition to accommodate the increasing demand for specializing training, especially in the labour market.

Among the various aspects that comprise the education field, learning is the one that requires the most careful treatment. The inherently complex domain lends itself to a wide range of forms and means, each with its techniques and quirks. Learning may happen at home, at work, or even on the go; thus, it is not limited to the classroom only nor confined to formal settings in general. Learning activities can (and should be) tailored around the individual. This personalization process is critical when targeting neurodiverse profiles or students with accessibility needs. Not only is content's form fundamental, but delivery and additional aids are required to make a learning experience impactful.

\textbf{Educational resources} are the primary means to help students in their learning journey. They can be of various kinds such as videos, interactive tutorials, pdf texts, images, podcasts and many others. Platforms such as \textit{Google Classroom}, \textit{Microsoft Teams}, or custom \textit{Moodle} deployments are common ways to deliver content in formal educational settings, but \textit{YouTube}, \textit{Wikipedia}, and platforms like \textit{Udemy} can reach a broader audience in informal environments. Furthermore, the increasing availability of Open Educational Resources (OER)\footnote{Open Educational Resources are public domain or open licensed educational resources \url{https://www.unesco.org/en/communication-information/open-solutions/open-educational-resources}} can help reduce teachers' material preparation and lower the accessibility barrier in terms of cost and material kind.

Learning is also crucial in the industry. The continuous technological disruptures are creating job positions in brand new fields\footnote{\url{https://www.linkedin.com/business/talent/blog/talent-strategy/linkedin-most-in-demand-hard-and-soft-skills}}. Entirely new hard skills are required to fit the openings. Moreover, soft skills are among the most sought-after skills because of their lower trainability and a slower changing pace\footnote{\url{https://www.linkedin.com/business/talent/blog/talent-acquisition/why-shell-pushes-hard-on-soft-skills}}. However, the offer seldom matches the demand. This mismatch led to the need for \textbf{upskilling} and \textbf{reskilling}. The former means teaching employees new, advanced and valuable skills to match the profile required for the next step in their current career path. The latter, instead, targets employees with a profile similar to the one required. It consists in teaching them adjacent skills and training them for their new, different job. Both of them are fundamental learning activities that take place in varied (and often dissimilar) environments.

From a student's perspective, effective teaching means \textbf{1-1 tutoring}\footnote{\url{https://www.eschoolnews.com/2022/07/26/ai-intelligent-bots/}}. It allows teachers to target specific misunderstandings and necessities with real-time feedback and explanations relevant to the student's experiences. On the other end of the spectrum, the most teacher-effective approach is the one-to-many lecture, where the teacher prepares the material upfront for being presented to a wide audience. One of this approach's main downsides is encouraging passive learning. A study by K.R. Koedinger et al. shows that the "Doer Effect" is a causal association between practice and learning outcomes and that practicing is six times more effective than reading \cite{DOER_EFFECT}. Another significant drawback lies in the motivational aspect. \textbf{Active learning} is more effective in learning outcomes and motivation than passive learning \cite{ACTIVE_LEARNING}. Despite that, a recent article on PNAS by L. Deslauriers et al. highlights a negative correlation between actual learning and the feeling of learning in the students \cite{ACTIVE_LEARNING_2}. However, \textbf{gamification} and serious games gained consensus as tools to motivate people to engage in beneficial activities, even if seen as unrewarding or tedious \cite{deterding2011gamification,GAMIFIED_APPLICATIONS,deterding2011game,BASSANELLI2022103657}. 

The design of a learning experience should take into account all of these different factors. On the other hand, \textbf{a design framework should not make assumptions about content type, form, delivery, and validation while still removing any obstacle between the teacher, the student, the environment, and the learning experience.} Individual coaching is rarely feasible due to poor scalability, whilst one-to-many general training is scalable but lacks individualization altogether. That is the gap the Personalized and Gamified eTutoring System - \Polyglot{} - aims to fill. Adaptive eTutoring systems support both teachers and students by combining the benefits of individualized delivery and manageability by leveraging software personalization, while gamification can be used to enhance motivation through personalized rewards or cooperative and competitive activities.

 The rest of the paper is structured as follows. Section \ref{sec:background} presents the needed background and the motivations that led us to realize \Polyglot{}. Section \ref{sec:polyglot} introduces its architecture and explains how it can help teachers create flexible learning and eTutoring experiences. AI planning and gamification mechanics needed for adaptive learning will also be discussed by the end of section \ref{sec:polyglot}. Section \ref{sec:demo} will describe a scenario where \Polyglot{} is used to assist teachers and students in introductory statistics for a data science course. Lastly, with Section \ref{sec:conclusion} we present a glimpse into future directions and possibilities enabled by the proposed solution.

\section{Background and Motivations}
\label{sec:background}
Adaptive learning technologies have gained traction over the last decade. Existing solutions have been successful in both domain-specific \cite{PRUSTY_ADAPTIVE} and institution-wise implementations. In 2015, the Colorado Technical University (CTU) reported that, following their \textit{Intellipath\texttrademark} adoption, the course pass rate rose sharply by 27\%, and the average grade and retention rate were also significantly affected. CogBooks\footnote{\url{https://www.cogbooks.com/}} conducted a pilot study with Arizona State University (ASU), resulting in findings similar to CTU.

Smart Sparrow\footnote{\url{https://www.smartsparrow.com/}} (recently acquired by the industry-leading Pearson) provides a user-friendly WYSIWYG content authoring tool to create interactive online experiences. Realizeit, Cerego and CogBooks create content for their adaptive platforms by partnering with institutions like ASU, the American Psychological Association, or directly with industrial partners and customers. Each platform also provides in-depth analytics on the students' performance, errors, time spent learning, and other related metrics. Those solutions solve the problem of individualization and verticality of linear learning paths but overlook a fundamental factor: familiarity with the tooling and the environment. Job candidates in software engineering, for example, are expected to be somewhat comfortable with industry-leading tools and methodologies. Therefore, not only delivering interactive exercises directly on those tools enables richer experiences but also better prepares the students for their future. High school electronics students, instead, may benefit from hands-on experience with physical devices like Arduinos\footnote{\url{https://www.arduino.cc/}}. Moreover, integrating education-ready tools like pi-top kits\footnote{pi-top produces high-quality educational kits for electronics and robotics based on the Raspberry Pi. \url{https://www.pi-top.com/}} or Swift Playgrounds\footnote{https://www.apple.com/swift/playgrounds/} allows access to already existing quality resources.

Novel interactive learning experiences can emerge even with the use of available technologies. Augmented and virtual reality are still emerging, but education and industry are already taking advantage of their benefits \cite{VR_AR_EDUCATION}. In the classroom, they are used to promote interactivity and generate engagement. In the industry, they found use in, among the others, remote maintenance on industrial devices, surgery training programs, aviation, and even military equipment\footnote{Since 2019 Microsoft and the US Army have collaborated on using HoloLens to enhance soldiers' situational awareness \url{https://news.microsoft.com/transform/u-s-army-to-use-hololens-technology-in-high-tech-headsets-for-soldiers/}}. Similarly, the use of voice user interfaces (like virtual assistants) may be a means to mimic a study companion that can ask questions, give feedback and help with misunderstandings. Adaptive educational technology should allow and encourage the creation of these varied activities to prepare the students for dynamic and changing environments where learning and adaptability are essential.

The main focus of current implementations of adaptive learning technologies is the content personalization engine. However, although being a crucial component of adaptive experiences, the \label{delivery} delivery of said personalized content happens through a custom-designed student platform. That may be a good fit for some activities, but it is a limiting factor for others, such as those in the aforementioned examples.

\begin{figure*}[t]
    \centering
    \includegraphics[width=.95\textwidth]{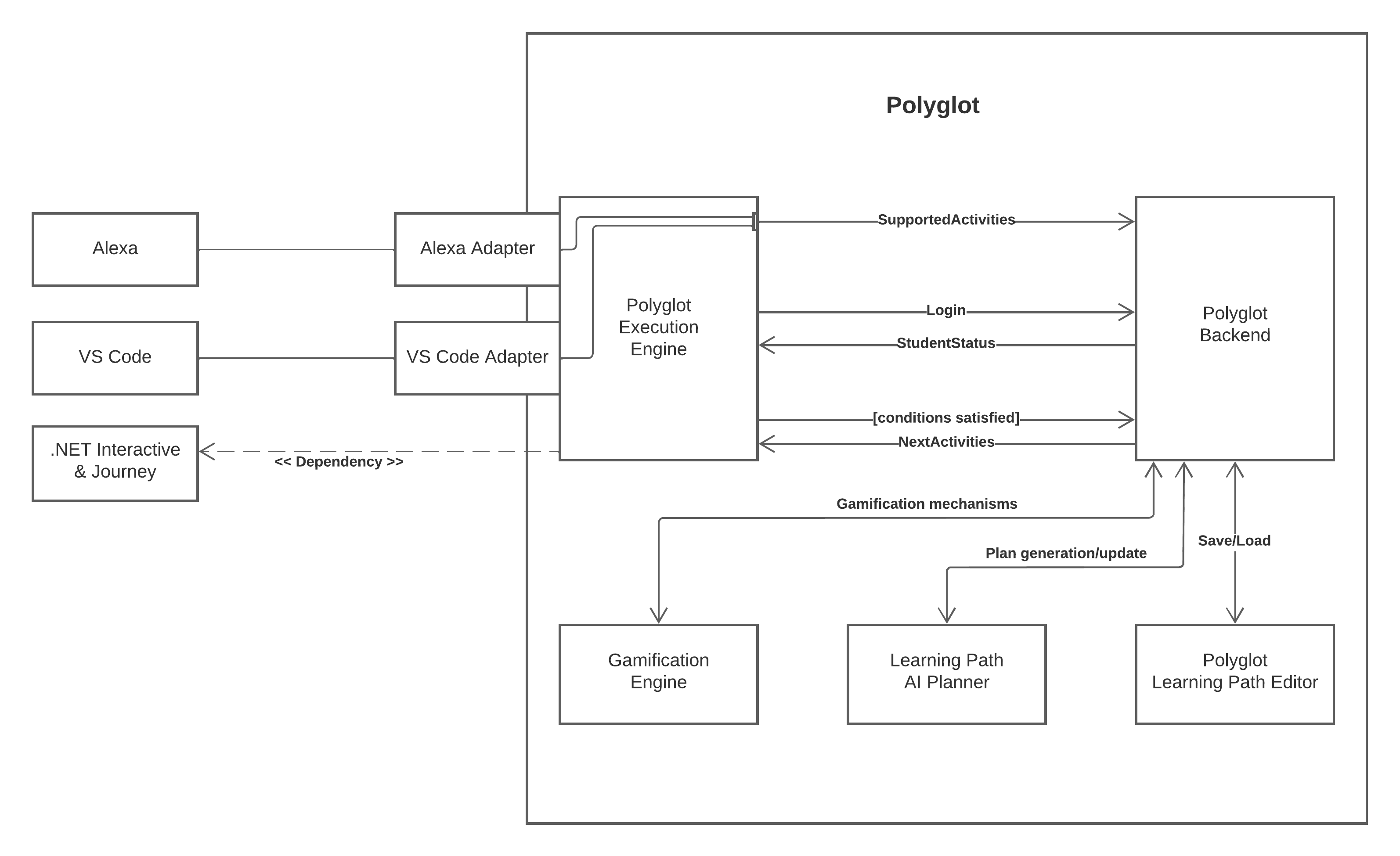}
    \caption[]{\Polyglot\ architectural overview.}
    \label{fig:polyglotArchitecture}
\end{figure*}

\section{\Polyglot}
\label{sec:polyglot}

\Polyglot\ aims to provide an open, content-agnostic and extensible framework (see Figure \ref{fig:polyglotArchitecture} for its architecture) for designing and consuming adaptive and gamified learning experiences. We imagine a student experience entirely tailored to their needs and choices. For example, we think students should be able to do some lessons and quizzes with Alexa, switch to VS Code to do some coding activities, and then move to another frontend (i.e., Moodle) to do something else, all without friction. That is why we exploit the flexibility of \texttt{.NET Interactive}\footnote{\url{https://github.com/dotnet/interactive}} in \Polyglot's execution engine. Students' interactions with external tools occur through \texttt{Adapters} that bind actions on the student frontend to \texttt{.NET Interactive} commands and bind \texttt{.NET Interactive} events to a supported output (e.g. audio for Alexa) with custom formatters.

The word "adaptive" in adaptive learning paths means being able to adjust the students' needs, like assigning exercises based on their previous responses or the capabilities of the platform they are on. This adaptation is allowed by the collaboration between the \texttt{Execution Engine} and the \texttt{Backend}. The former handles the students' submissions and validates their responses, while the latter uses the results of the validation phase to assign the next activity in the learning path.

\begin{figure*}[ht]
    \centering
    \includegraphics[width=\textwidth]{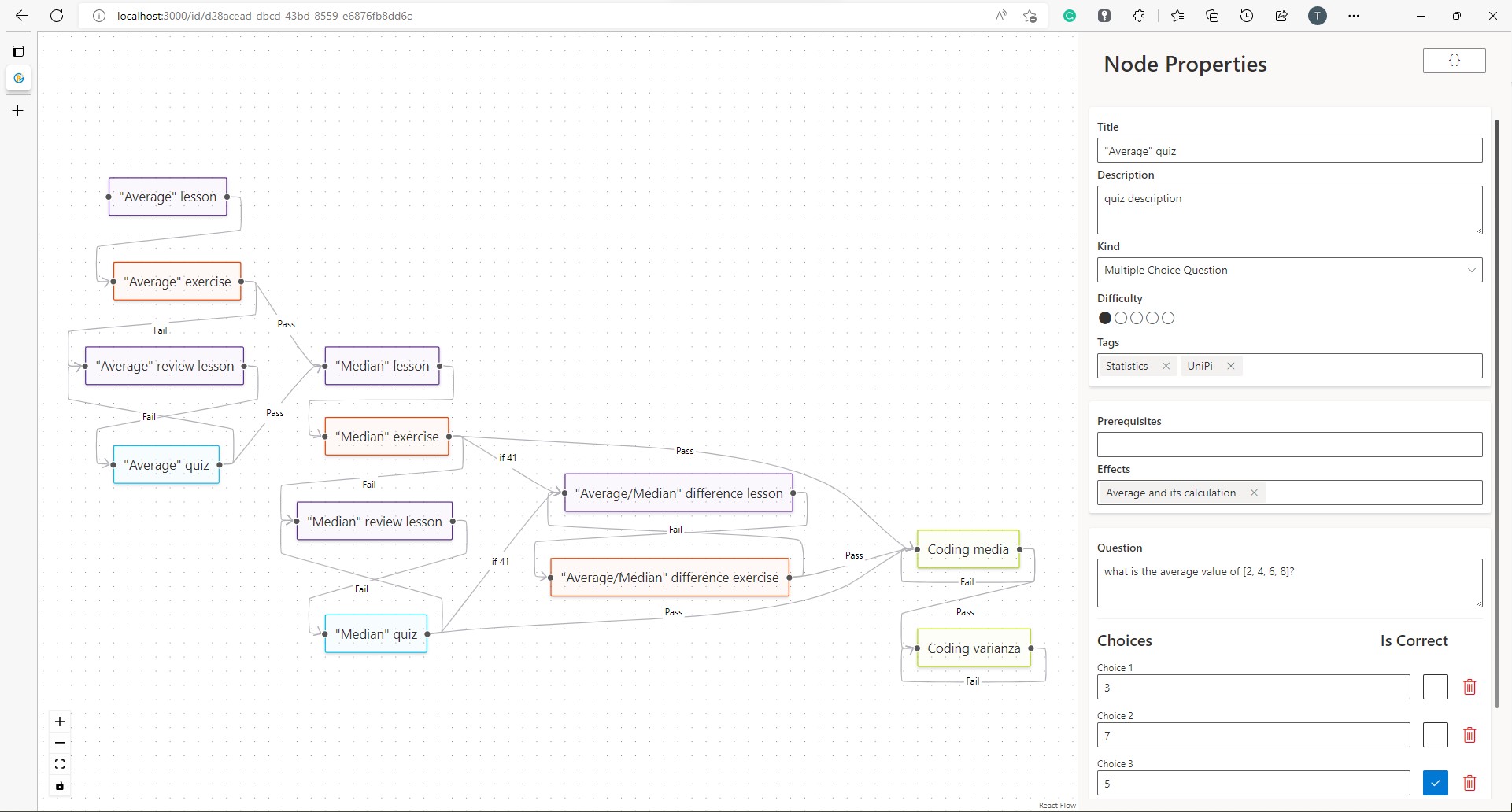}
    \caption[]{Teacher design tool for learning fragments.}
    \label{fig:teacherFrontend}
\end{figure*}

The two most important aspects of a tutoring platform are \textit{simplicity in the content creation} process and \textit{content availability} to students for consumption and teachers for reuse. 

Designing and realizing learning fragments for such a variegated learning experience might seem a daunting task at first. That is why \Polyglot\ includes a \texttt{Learning Path Editor} heavily focused on the teachers' experience (see Figure \ref{fig:teacherFrontend}). Its graph-like, visual-editing capabilities and integration with the rest of the platform allow the creation of ready-to-use learning fragments with ease, thanks to the abstractions provided. The teacher design tool is composed of two main elements: the \textit{drawing area} and the \textit{properties panel}. The drawing area is the core of the visual editing experience. Nodes can be added, connected, and rearranged with simple clicks or drag-and-drop interactions. Nodes and edges themselves can be interactive components that augment the visual editing experience. The properties panel, instead, provides fine-tuning tools to define the activities and the links between them. 

To define the learning fragment, a teacher can create new activities with a right-click on the canvas, change their type and parameters from the properties panel, and connect them by dragging from the handle on the source node to the handle on the destination node. By clicking on an edge, the teacher can edit the link condition by choosing from a list of existing abstractions or by writing their own validation code.

The tool is implemented using web technologies with React\footnote{\url{https://reactjs.org/}} and TypeScript\footnote{\url{https://www.typescriptlang.org/}} with a particular focus on flexibility and extensibility. Its design allows it to be part of a more comprehensive learning management tool. The drawing area and the properties panel are different abstract views that operate on the same data differently. Both components manipulate the same PolyglotElements (i.e. nodes and edges) by working on the shared state via Zustand\footnote{Zustand is a lightweight, unopinionated state-management library for React \url{https://github.com/pmndrs/zustand}} actions.

The \textbf{adaptation process} necessary for effective eTutoring relies on abstract activities: activities defined only in terms of their goals and not on concrete exercises the students have to solve. Those abstract activities are then replaced during the runtime refinement phase with the most suitable composition of existing fragments that satisfies the goal decided by the AI planner. The latter exploits the automatic generation of learning paths by adapting an existing approach based on AI planning presented in \cite{BertoliICWS09}. The chosen fragments may have other abstract activities, so further refinement stages may be needed.

\textbf{Gamification} means creating a game narrative that guides players through increasingly complex challenges, keeping them engaged with social activities such as group work or competitions. It means providing immediate feedback (as expected from a game-like environment) and students taking autonomous choices to progress down the individually decided path. Gamification is not an add-on. Instead, gamification mechanics are fundamental to the learning path personalization process in two ways. Not only do they keep the students engaged, but they can also be used as tools to gain insight into the student's behaviour from a different perspective and thus help generate a more personalized and engaging learning path. In order to increase engagement, the gamification mechanics must be calibrated according to the underlying activities. That is why \textbf{gamification fragments} may be composed by the AI planner similarly to learning fragments.

\section{The {\Polyglot} Demonstration}
\label{sec:demo}

The chosen scenario is part of an introductory statistics course that fits a broader data science program. Its goal is for the student to understand the concepts of average, median, and the difference between them. We have identified four kinds of concrete learning activities needed to define the scenario:
\begin{itemize}
    \item lessons
    \item close-ended questions
    \item quizzes
    \item coding activities
\end{itemize}
We want students to first learn about the average, then the median, and then how to code them. We also want students to \textit{understand} the average, the median, and the difference between them, so the fragment includes a review lesson and an additional quiz in case the student fails the close-ended question. Once the student has completed the "average" section, they move to the median, where they will face the same pattern for recovery used for the average. Suppose that the closed-ended question requires the student to calculate the median. Since understanding the difference is part of the goal, if the student answers with the average value instead of the median, they should be given a particular review lesson that highlights the differences instead of a regular review lesson. All those different alternative paths merge before the coding activities. While the other exercises have a known correct answer, coding activities do not. Furthermore, to be considered correct, a program may not only be required to have a correct output (e.g. through unit tests) but also to have some properties on the source code itself (e.g. low cyclomatic complexity\footnote{\url{ https://en.wikipedia.org/wiki/Cyclomatic_complexity}} or proper encapsulation\footnote{\url{https://en.wikipedia.org/wiki/Encapsulation_(computer_programming)}}) or on its execution (e.g. time or memory constraints). As evidenced by the previous example, fragments must not pose any limit on the relationship between inner activities. The complete fragment's implementation is shown in Figure \ref{fig:teacherFrontend} using the \Polyglot\ \texttt{Learning Path Editor}.

\begin{figure}[ht]
    \centering
    \includegraphics[width=.45\textwidth]{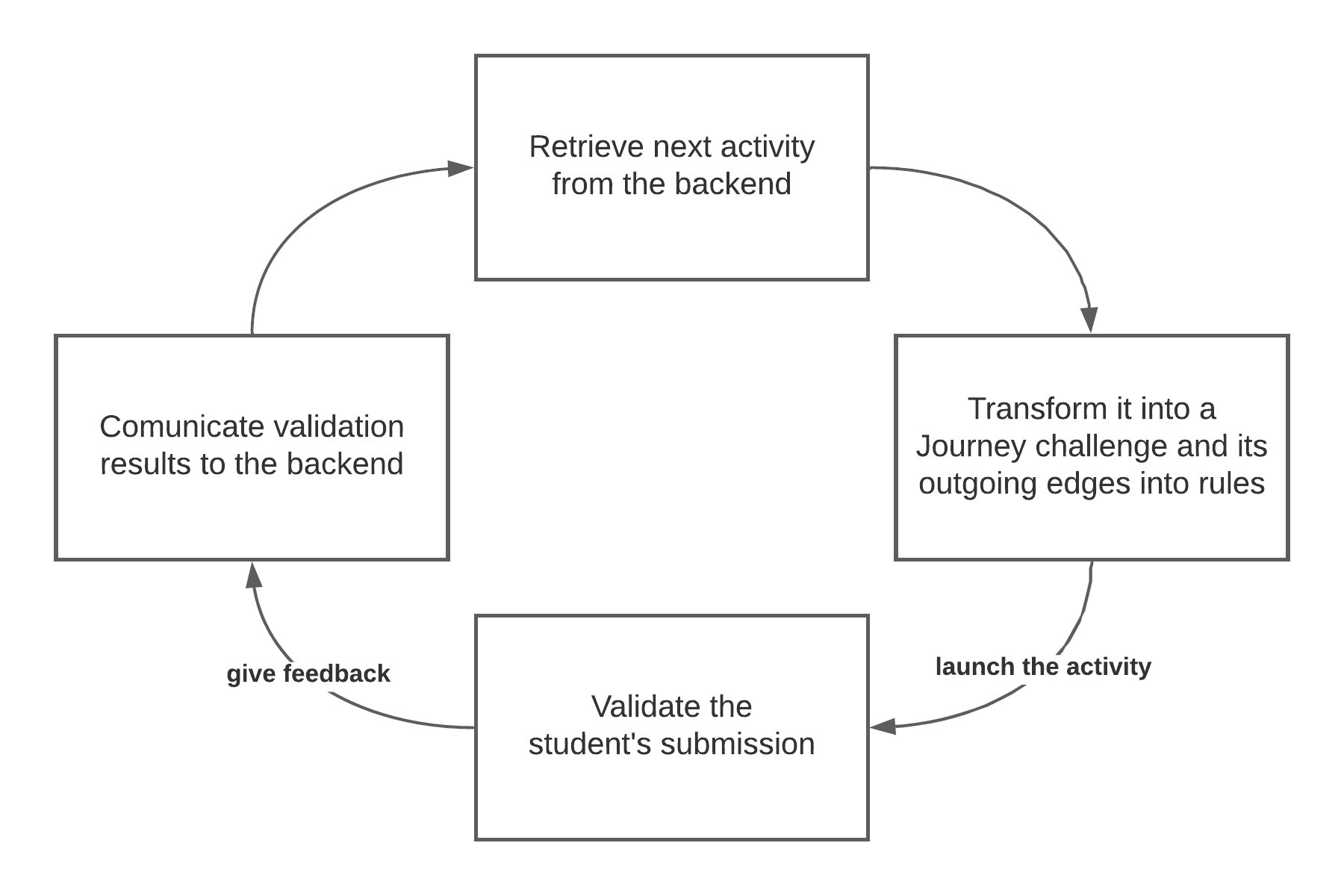}
    \caption[]{Execution Engine Lifecycle.}
    \label{fig:engineLifecycle}
\end{figure}

\begin{figure}[ht]
    \includegraphics[width=0.3\textwidth]{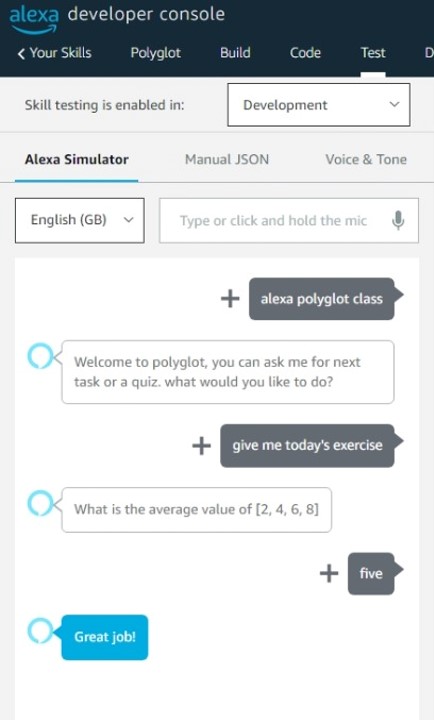}
    \captionof{figure}{Alexa frontend interaction.}
    \label{fig:alexaFrontend}
\end{figure}

\begin{figure}[ht]
    \includegraphics[width=0.4\textwidth]{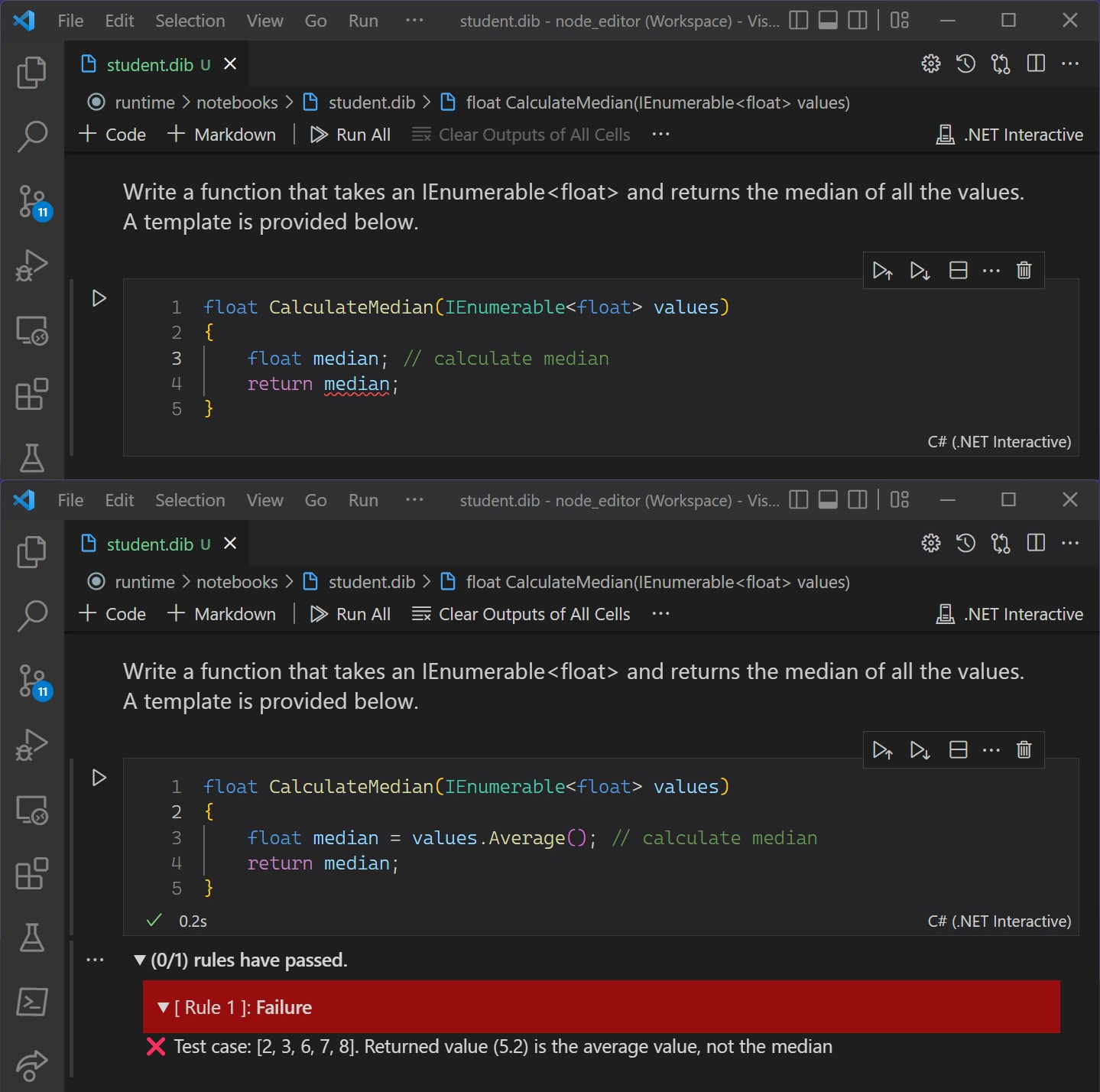}
    \captionof{figure}{VS Code frontend interaction.}
    \label{fig:vscodeFrontend}
\end{figure}

The student can consume learning fragments thanks to \Polyglot's \texttt{Execution Engine}. This component constitutes the core of the student experience. It is responsible for handling user interactions and validating students' submissions. Its straightforward lifecycle (see figure \ref{fig:engineLifecycle}) is designed to exploit \texttt{Journey}\footnote{Journey is a .NET Interactive library for defining open-ended graphs of challenges}'s strengths by using it in a semi-controlled fashion. When the assigned activity and its outgoing edges are retrieved from the backend, they undergo a transformation phase where the activity is converted into a \texttt{Journey} challenge and edges into \texttt{Journey} rules. This process is essential because it allows running the validation code mentioned previously. Once converted, the challenge is ready to be launched. The activity is "shown" to the student in their frontend, thanks to \texttt{.NET Interactive} formatters. When the active frontend adapter produces an event, \texttt{Journey} intercepts it and runs the registered rules against the student submission. \Polyglot{} then sends the result of this validation phase back to the backend that will use this additional knowledge to suggest the next suitable activity. At the moment of writing, the students can execute the learning fragment with Alexa (see Figure \ref{fig:alexaFrontend}) or VS Code for coding activities (see Figure \ref{fig:vscodeFrontend}).


\section{Conclusions and Future Directions}
\label{sec:conclusion}
In this paper we presented an innovative eTutoring platform characterized by its adaptivity and gamification mechanics. An overview of its architecture has been presented together with a practical demonstration to deepen how processes are linked and how the framework would handle the activities lifecycle. The work done and presented here has been a good experiment to set-up a set of research directions that we highlight in the following.

Adaptive education still remains an open research topic at the intersection of several different domains like cognitive sciences and informatics. We have identified some directions to explore in the future:

\begin{itemize}
    \item \textbf{Abstract activities} are at the core of our formalization. Extending \Polyglot{} to include them is our highest priority. To do that, we need to include them in the editor, design a tool for defining goals, and integrate the AI planner to include the runtime refinement and harness the power of abstract activities.
    \item Runtime refinement is one of the most powerful concepts in the entire framework. However, it is most effective only if there is a substantial amount of fragments to choose from. We think that creating an \textbf{open fragment database} would benefit students (with better refinement and high-quality educational content) and teachers who can focus on the big picture and let the platform adjust the details.
    \item To further extend the previous point, \textbf{leveraging open educational resources with NLP techniques} for automatic content annotation would significantly contribute to the database. This task is one of the main goals of Project Encore\footnote{\url{https://grial.usal.es/encore}}, an ERASMUS\texttt{+} project that involves multiple universities and companies.
    \item Allowing students to use \textbf{arbitrary frontends} for their tasks is indeed challenging. This support is still experimental and needs more work to ensure the quality of the learning experience. Furthermore, not all frontends are suitable for all kinds of tasks: for example, using Alexa to do a coding exercise might not be the best idea. We need to further define what "suitable" means and implement something similar to a content negotiation mechanism.
    \item To easily define the fragments, \Polyglot{} provides some basic abstractions for connections such as "Pass/Fail". As explained previously, this abstraction may be correct for some exercises but not applicable to others. We want to further explore this topic to enhance the correctness of existing fragments, \textbf{validate connections} to prevent the creation of erroneous links, and define other composable abstractions.
    \item We plan to add \textbf{support for existing learning management systems}\footnote{\url{https://en.wikipedia.org/wiki/Learning_management_system}} (e.g. Moodle LMS) in the same modular fashion as we are doing with students' frontends. This extension would allow incremental integration of parts of \Polyglot{} within existing learning infrastructures and increase the adoption of adaptive learning technologies.
    \item The combined evolution of the engagement and learning systems is still an open research problem. Gamification mechanics are effective only when they fit the learning path perfectly. Usually, it is the gamification designer's duty to design the game narrative and the other game elements to achieve that fit. However, the power of adaptive learning paths comes from the runtime refinement; therefore, paths are not available upfront. The engagement system must then evolve through an \textbf{automatic calibration} phase to the mechanics to the underlying activities.
    \item The learning path planning is crucial for effective refinement. We aim to enhance the planner by leveraging \textbf{machine learning techniques using learners' behavioural data}. Our goal is to improve contextual activity suggestions (and hopefully feedback), considering non-standard parameters like the capabilities of the interface the student is using. 
\end{itemize}

\paragraph{}

\section*{\uppercase{Acknowledgements}}
This work is supported by the Erasmus+ 101055893 project , named "ENCORE -ENriching Circular use of OeR for Education", funded by the European Commission.

\bibliographystyle{ACM-Reference-Format}
\bibliography{main}

\end{document}